.

# SOCIO-ECONOMIC MODELS OF DIVORCES IN DIFFERENT SOCIETIES

**Andrzej Jarynowski[1], Marta Klis[2]**

[1]Department of Complex Systems, Smoluchowski Institute of Physics, Jagiellonian University
ul. Reymonta 4, 30-059 Kraków,
[2]Faculty of History, Nicolaus Copernicus University
ul. Bojarskiego 1, 87-100 Toruń
andrzej.jarynowski@uj.edu.pl, martaklis@yahoo.co.uk

**ABSTRACT**

Population dynamic of getting divorced depends on many global factors, including social norms, economy, law or demographics as well as individual factors like the level of interpersonal or problem-solving skills of the spouses. We sought to find such a relationship incorporating only quantitative variables and test theoretical model considering phase transition between coupling (pairs) and free (single) preferential states as a function of social and economic . The analyzed data has been collected by UN across almost all the countries since 1948. Our first approach is followed by Bouchaud's model of social network of opinions [1], which works well with dynamics of fertility rates in postwar Europe. Unfortunately, we postulate that this pure sociological and pure economic approach fail in general. Thus, we did some observation about why it went wrong and where economy (*e. g.* Poland) or law (*e. g.* Portugal) has bigger impact on getting divorce than social pressure.

## "FREEZING A LOT, TYING A KNOT": MODEL ASSUMPTIONS

Let assume that marital status has its mechanistic equivalent in the state of matter observed in nature. People can be in two states: free (single) and coupled (married), and are allowed to transmute between these states. Let introduce analogy to phase transition from physics, described in general as a function of temperature and pressure. We can observed similar patterns in people's behavior where social and economic freedoms have impact on numbers of weddings or divorces "Fig. 1". This theory is a framework to all analysis.

There are also weak points of such approach. Sensitivity to local phenomena, called idiosyncrasy, makes difficult the comparative analysis of different countries. Demographic structure of society changes in time, and is specific for each country.

We incorporate to the model only divorce numbers or rates in different countries. We tried to observe and categorize social and economical factors, which can contribute to peoples decision to get divorce. Economic pressure or freedom of people living in different regions can be measured directly (*e.g.* salaries, average space per person and other indicators easily accessible from statistical comparative reports). Unfortunately, there is no direct measure of social freedom or pressure. We assumed that observation of dynamic (particularly growth) of divorce rate in many countries can give us insights how the social pressure collapsed in last 50 years. There are some sociological theories [2] of observed privatization of religion and social life at all, which concluded with stopping listening to church or social authorities (in all of biggest world's religion divorces are in particular forbidden). Nonlinearity in dynamics of divorce rate can be produced by feedback loops between the social norm's change and people's behavior.

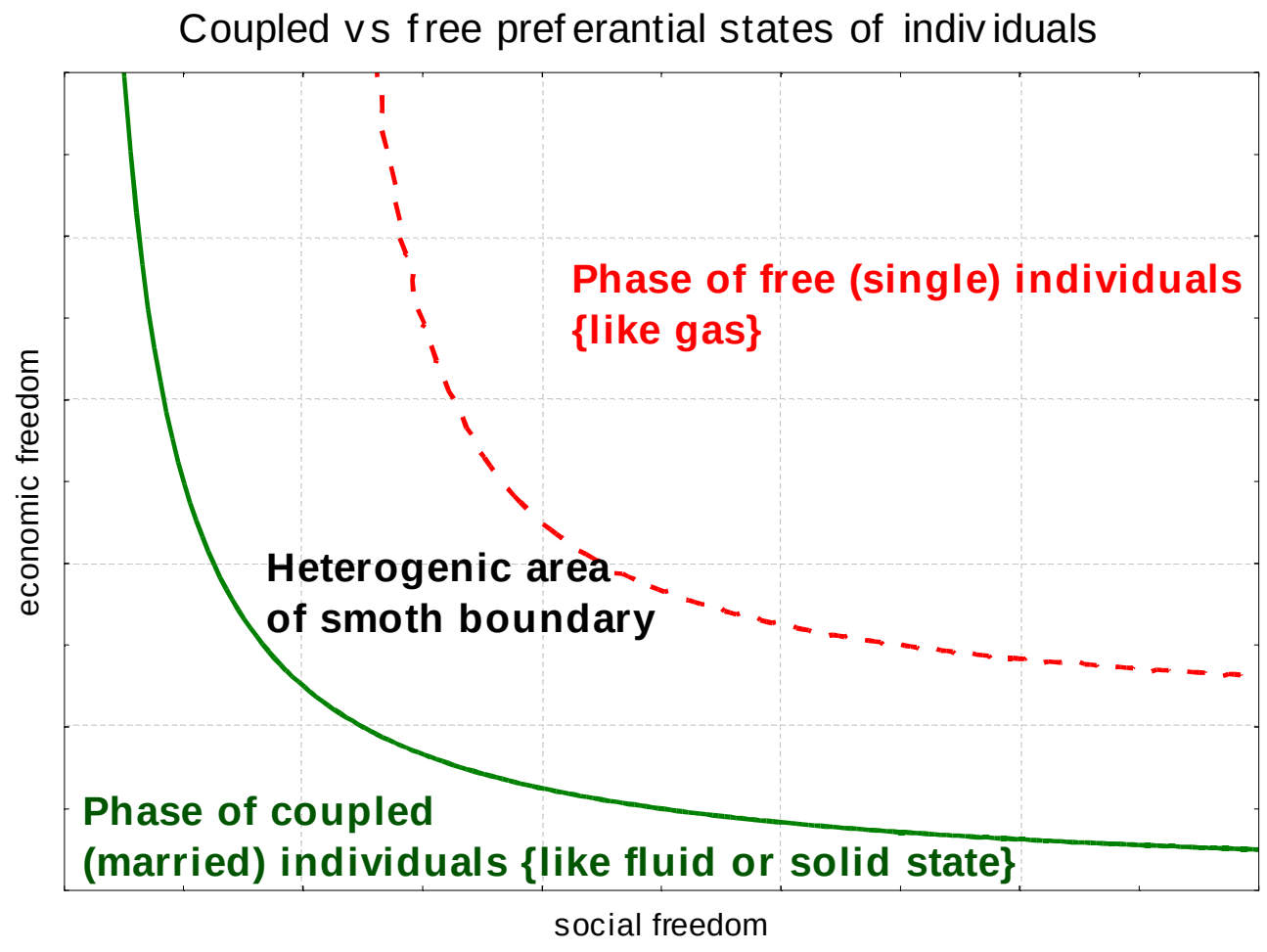

Figure 1. Framework of marriage-divorce model

Let define that state X of every person in population can be in two states: 1– single or -1 – coupled. Dynamics of change of state of person i can be phenomenological "Fig. 1" described as:

$$E_i(t)\,S_i(t) = T_i(t) \tag{1}$$

Where: $E_i(t)$ is a economic freedom which can be for *e.g.* linear reaction on salary;

$S_i(t)$ is a social freedom which depend on many factors like social norm level, susceptibility on social pressure or influence of other people;

$T_i(t)$ is a total freedom and indicate if person should stay or change state so it has to be compared with threshold value for every individual.

In this paper both economical freedom and social freedom are estimated separately.

**NORM BREAKING RULE AND SOCIAL CHANGE [1]**

Collective effects induced by imitation and social pressure were analyzed many times for different areas of social life. Shifts of opinions can occur either abruptly or continuously, depending on the importance of herding effects, sometimes called Zeitgeist in literature. Particularly interesting and generic "Random Field Ising Model" (RFIM) has been primarily successfully proposed to model of hysteresis loops in random magnets. The model was easily translated in order to represent a binary decision situation under social pressure, influenced by some global information [3]. Hamiltonian of such a system is calculated from neighborhood of every agent. In the first approach lattice network is proposed, but other network structure can be applied like: "Small World [4]", scale free or standard random networks.

In general, social freedom can be described as (when economic freedom is negligible):

$$T_i(t) = S_i(t) = \phi_i + F(t) + \sum_{j \in D_i} J_{ij} X_i \tag{2}$$

Where: $\Phi_i$ is an individual susceptibility on social pressure;

$F_i(t)$ is a power of social norm and can be understand as a external field;

$J_{ij}$ is a power of relation between individuals *i* and *j* and for simplification can be unified to one value *J* for every pair;

$D_i$ is a neighborhood of individual *i*.

Transmission driven by social pressure (freedom) can have two main realizations:

a) domination of social norm (J~0) - polarization of opinion changes smoothly with time-dependent social pressure;

b) domination of imitation effect (J>>0) - decisions change dramatically with a certain threshold value.

Mean field approach allow to investigate “Eq. (2)” for well-connected societies. In our analysis, we investigate an increase of divorces number, or rather the scaling between the height *h* of the peak respectively to its width *w*. The speed of change generically peaks at a certain time; the main prediction is a scaling relation $h \sim w^{-k}$. In the mean field solution, the parameter *k* when close to 1 corresponds to model a) with a simple pressure function, but close to 2/3 calls for model b) with imitation. Bouchaud checked two sets of data human behavior changed because of “social revolution” and there are compatible with this prediction, with *k*~0.62 for birth rates, *k*~0.71 for cell phones. Unfortunately, this pure sociological model is not with agreement with divorce data.

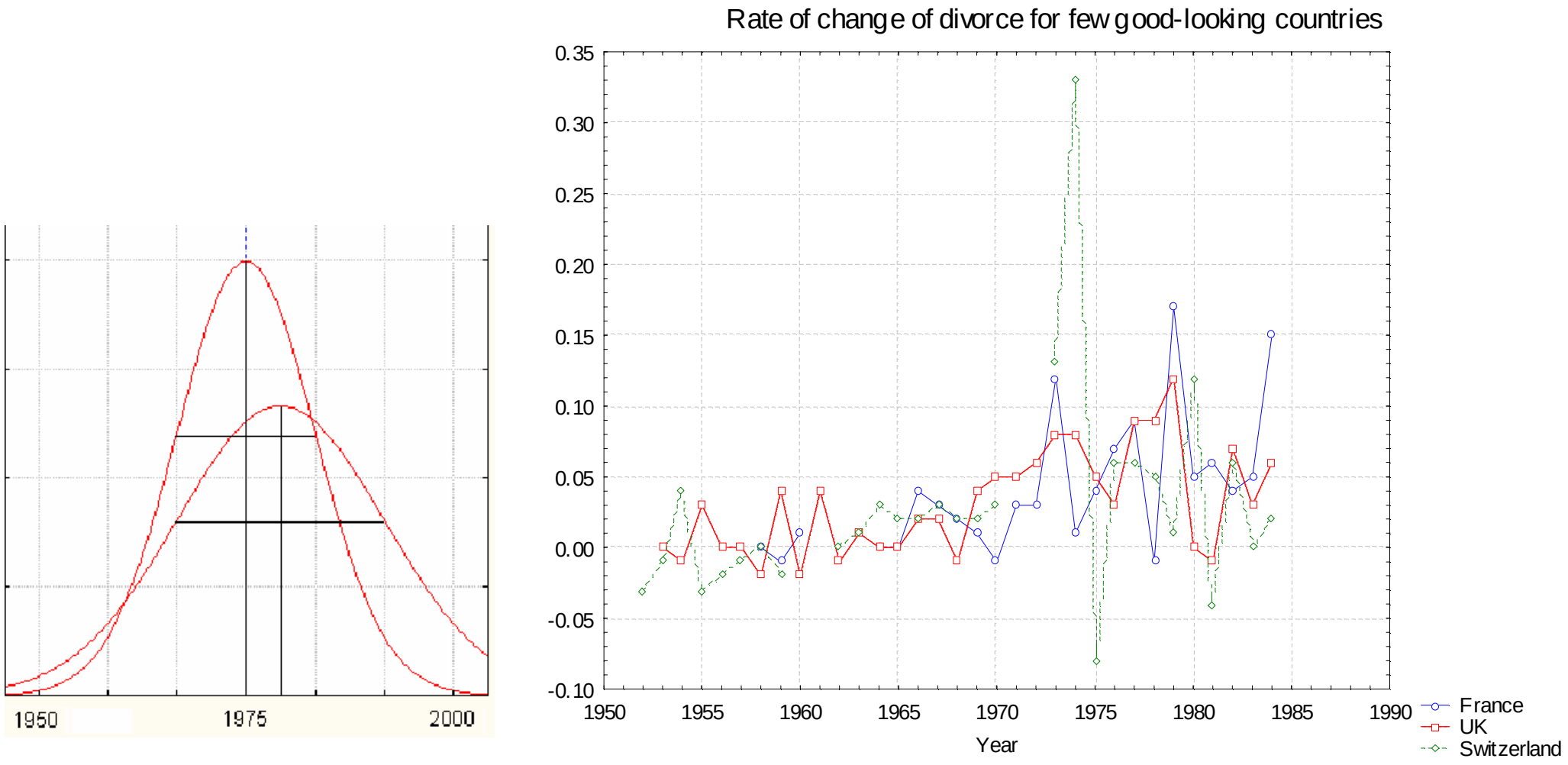

Figure 2. Ideal fitting procedure (left) and some real trajectories of divorce rate (right)

### SOCIAL CHANGE MODEL: TECHNICS

In an idealized situation, if the increase of the divorce rate were Gaussian “Fig. 2-left”, our calculation would be easy (as Bouchaud postulated). Real trajectories “Fig. 2-right” for different countries are much more noisy (not only social norm change influenced divorce rates and political, economic or low impacts composed background noise). There are two estimators of w and *h* for each country: Inverse variance method [5], based on the calculation of volatility of data, were firstly introduced to estimate *w* and *h* “Fig. 3-left”. Bayesian method [6] uses different cumulated percentiles of data distribution and integrates over all possible cases “Fig. 3-right”. Second (Bayesian) estimator seems to be more mathematically correct but more naive approach (inverse variance) fits better. Generally, both estimators do not fit the data successfully. So that, we conclude that pure sociological approach do not describe the marriage-divorce phenomena.

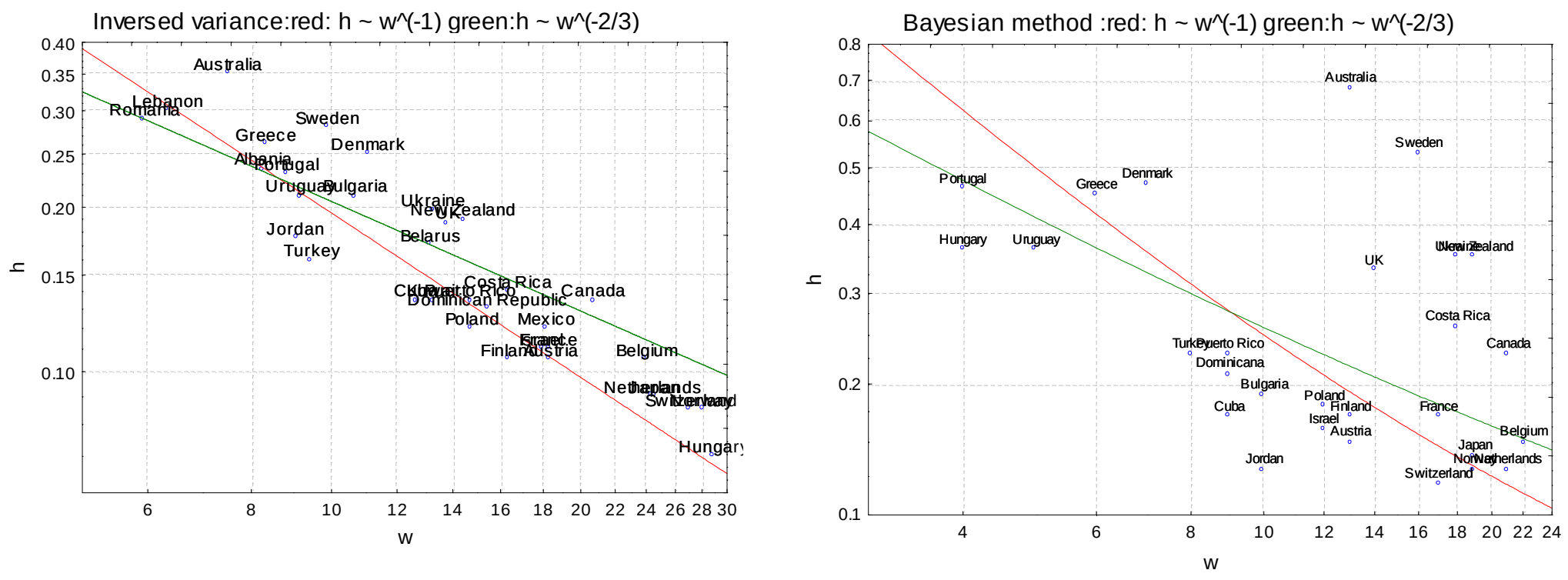

Figure 3.Inverse variance (left) and Bayesian method (right)

## PURE ECONOMIC APPROACH

We run regressions on Polish data, where independent variable is an average salary in region and dependent one is the rate of divorces. We assume, that social freedom is constant in the whole country, therefore total freedom is defined as a linear response to income of individual “Eq. 3”. Unfortunately, there is no significant relation “Table 1”, although some regularity is observed. *E.g.* the richest state (Warsaw) has three times bigger rate than the poorest one (Podkarpacie – Polish Galicia). Other observed decreases of divorce number can be locally explained by temporal national economy state.

$$T_i(t) = E_i(t) = b_{income}\, Income_i(t) + b_{free\ el.} \qquad (3)$$

Table 1. Regression test for divorce rates explained by income grouped by Polish voivodeships for years 2011 [7] and 2010 [8]. P-Value of test of explanatory variable is above any reasonable confidence, so relation is insignificant.

| N=34 | b | err. b | t(32) | p-Value |
|---|---|---|---|---|
| Free el. | 1.86 | 0.64 | 2.69 | 0.01 |
| Income | 0.00 | 0.00 | 0.28 | 0.78 |

## CONCLUSIONS AND FUTURE ANALYSIES

Measuring social and economic freedom at once is extremely difficult. We proposed indirectly pure social or direct pure economic estimates, but both do not explain phenomena separately. Norm changes, easily explained by Bouchaud's model to cells sells or to decision on having children is not seen explicit in divorces dynamics. One of the biggest factor of mismatching methods is demography (especially of postwar Europe). Number of peoples was permanently growing with different speed. Age-structures of societies were also changing over time. Rich countries gained big amount of immigrants in waves. Cell phones sale dynamics have reached saturation level just in few years, for that reason the change of demographic structure do not disturbed this process. On the other hand, birth rate observed for the last 50 years, was operationally defined as a fertility rate. This definition allowed to avoid any problems of demographic structure, because of normalization of births to number of woman in fertile age. There is no perfect way to exclude age-structure of societies in order to obtain “clean” divorce rates. Moreover, bigger problems appear if law of politics changes. Carnation revolution in Portugal or Political transformation in Poland have brought crucial influence on divorce dynamics “Fig. 4”. Salazar's regime breakdown concluded with creating new, more liberal law, which allowed people to divorce, because earlier it was almost impossible, and divorces grew dramatically. On the other hand, in Poland when economic and political situation were difficult, number of divorces fall down, because during that time people were looking for a support in their partners.

Table 2. Regression test for divorce rates explained by support to conservative party ”Pis” grouped by Polish voivodeships during parliamentary elections in 2011 [9]. P-Value of test of explanatory variable is below any reasonable confidence, so relation is significant and negative.

| N=17 | b | err. b | t(15) | p-Value |
|---|---|---|---|---|
| Free el. | 3.32 | 0.25 | 13.32 | 0.00 |
| “PiS” supporters | -0.04 | 0.01 | -5.31 | 0.00 |

We have in mind some other estimators like voting percentage (liberal *vs* conservative) which can structured in spatial-temporal way as economic factor. We have even shown relation between voting patterns and divorce rates on level of Polish voivodeships (percentage support of conservative party in election 2011 “PiS” has significant negative impact on divorce rate), but supporting political parties is socio-economic process and cannot be factorized on separated social and economic indicators.

We are planning to study marriage-divorces mechanism on spatial-temporal networks. There is possibility to go even further, and look into transition single-in relation for no-registered couples. That information could be collected via Facebook or surveys. We can measure directly economic freedom (usually as an average salary in region in time series or sometimes even on individual level) but social pressure must be defined. In our work we used Bouchaud method that let us to obtain social factor irrespective to time series. We had to look at 40-50 years time interval to deduce value of social pressure change. those disadvantages of Bouchaud's methods has just a small impact on comparison with non-compliance between theory and data “Fig. 3”. This global approach fails. We could look then on dynamics of society networks where nodes would be individuals and links marriages or relations. Those individuals would be exposed on local socio-economic redefined factors and agent-based modeling would be applied to compare with real processes.

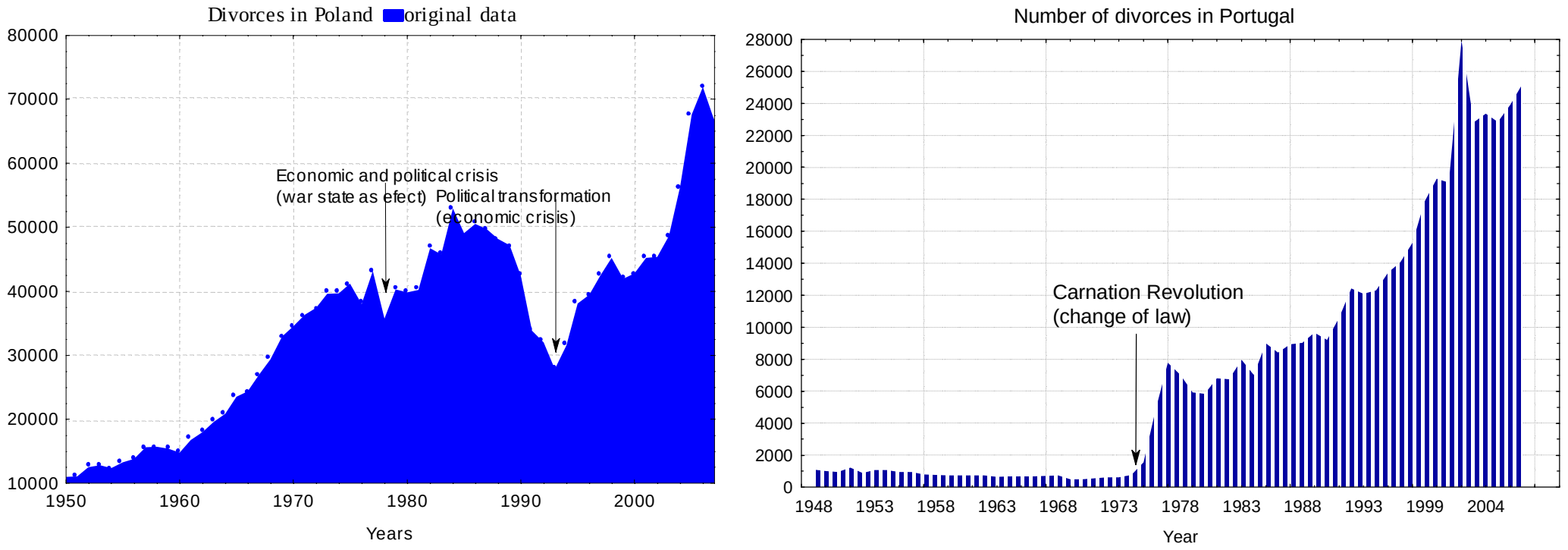

Figure 4. Divorces in time with some events. Poland (left) and Portugal (right)